# Universal relaxor polarization in $Pb(Mg_{1/3}Nb_{2/3})O_3$ and related materials

A. A. Bokov, and Z.-G. Ye*
*Department of Chemistry, Simon Fraser University, Burnaby, BC, V5A 1S6, Canada*

The dielectric permittivity $\varepsilon$ at frequencies from ($10^{-1}$ –$10^{-5}$) Hz to $10^5$ Hz is studied in perovskite (1-$x$)$Pb(Mg_{1/3}Nb_{2/3})O_3$ – $x$$PbTiO_3$ relaxor ferroelectric ceramics of different compositions $x$ = 0.35, 0.25 and 0, which exhibit, below the temperature of the diffuse $\varepsilon(T)$ maximum $T_m$, a tetragonal ferroelectric, a rhombohedral ferroelectric and a nonergodic relaxor phase, respectively. The universal relaxor dispersion previously observed at temperatures near and above $T_m$ in the ceramics of $x$=0.25, is also found to exist in other compositions. This dispersion is described by the fractional power dependence of the real and imaginary parts of susceptibility on frequency, $\chi'_U(f) \propto \chi''_U(f) \propto f^{n-1}$. The real part of the universal relaxor susceptibility $\chi'_U$ is only a comparatively small fraction of the total permittivity $\varepsilon'$, but $\chi''_U$ is the dominant contribution to the losses in a wide frequency-temperature range above $T_m$. In the high-temperature phase a divergent temperature behavior is observed, $\chi'_U(T) \propto (T-T_0)^{-\gamma}$ and $\chi''_U(T) \propto (T-T_0)^{-\gamma}$, with $T_0 < T_m$ and $\gamma \cong 2$, for all the three compositions studied. The universal relaxor susceptibility is attributed to the polarization of polar nanoregions, which are inherent in the relaxor ferroelectrics. A microscopic model of this polarization is proposed, according to which the dipole moments of some ('free') unit cells inside polar nanoregion can freely choose several different directions, while the direction of the total moment of the nanoregion remains the same. The ensemble of interacting polar nanoregions is described in terms of a standard spherical model, which predicts the quadratic divergence of susceptibility above the critical temperature, in agreement with the experimental results.

PACS numbers: 77.22.Gm; 77.22.-d; 77.80-e; 64.60.Fr; 77.84.Dy

## I. INTRODUCTION

The structure and properties of relaxor ferroelectrics are nowadays a fascinating puzzle that has attracted a great deal of theoretical and experimental work.[1] The appropriate description of the temperature dependences of the dielectric permittivity with broad peak and extremely high value, typically observed in these materials, is a necessary and important step toward the understanding of the problem. In the studies of the dielectric response of $0.75Pb(Mg_{1/3}Nb_{2/3})O_3$ – $0.25PbTiO_3$ (PMNT75/25) relaxor ferroelectric with the perovskite-type structure, we recently found that in the temperature range above the ferroelectric phase transition, the most significant contributions to the dielectric permittivity ($\varepsilon = \varepsilon' - i\varepsilon''$) come from two distinct mechanisms, called the "conventional relaxor polarization" and the "universal relaxor polarization", respectively.[2,3] The first contribution gives rise to the diffused maximum in the temperature dependence of the real part of the permittivity $\varepsilon'$ and amounts to about 80-90 % of the total $\varepsilon'$ value. The real part of the "universal relaxor" susceptibility, which results from the 2nd contribution and provides the remaining 10-20 % of the total $\varepsilon'$, was found to follow the critical relation

$$\chi'_U = C'_\chi (T-T_0)^{-\gamma} \quad \text{at } T > T_0, \qquad (1)$$

where $C'_\chi$ and $T_0$ are the parameters slightly dependent on frequency and the critical exponent $\gamma$ equals 2. When temperature approaches $T_0$ from above, this critical relation is violated and $\chi'_U(T)$ dependence becomes rounded. As a consequence, the sum of the two contributions, conventional plus universal, remains non-critical.

* Corresponding author: FAX: (604)-291-3765. Electronic address: zye@sfu.ca

The second remarkable feature of PMNT75/25 is the Curie-Weiss law for the total permittivity observed in the low-temperature phase:[4]

$$\varepsilon' = C(T_C - T)^{-1} \text{ at } T < T_C . \quad (2)$$

The Curie-Weiss temperature $T_C$ is approximately equal to the critical temperature $T_0$ and is several degrees below the temperature of the diffused $\varepsilon'(T)$ maximum ($T_m$).

The validity of the Curie-Weiss law (2) below the Curie point implies that the spontaneous transition from the relaxor to a normal (long-range ordered) ferroelectric (FE) phase, which is known to exist in 0.75PMN - 0.25PT at $T_C$, must be a second-order (or a first-order close to the tricritical point) phase transition.[4]

Several issues still remain to be solved with regards to the key features of the universal relaxor susceptibility, in particular,

i) whether or not the critical behavior (1) is related to the spontaneous transition from the relaxor to a normal ferroelectric phase, or it can be observed in relaxors without such a transition;
ii) whether or not the critical behavior (1) is connected with the fact that the spontaneous transition to the ferroelectric phase manifests itself as a second-order (or nearly) phase transition;
iii) what the polarization mechanism responsible for the universal relaxor susceptibility is and what the nature of the critical behavior (1) is.

The aim of the present work is to elucidate these issues. We have observed the "universal relaxor" dielectric behavior both in $0.65Pb(Mg_{1/3}Nb_{2/3})O_3$ – $0.35PbTiO_3$ (PMNT65/35) ceramics, which undergoes a phase transition from the relaxor to the normal ferroelectric phase without obeying the Curie-Weiss law at $T < T_C$, and in pure $Pb(Mg_{1/3}Nb_{2/3})O_3$ (PMN) ceramics where a normal ferroelectric phase does not exist at all and the material remains in the nonergodic relaxor state. It is found that the universal relaxor susceptibility shows the critical quadratic behavior in both cases. In order to explain the unusual critical behavior described by Eq. (1), we propose a microscopic model for the polarization of the reorientable polar nanoregions (PNR), which inherently exist in relaxors and were suggested to be the origin of the universal relaxor susceptibility.[3] This model allows us to describe the relaxor system in terms of the standard spherical model. It leads to a diverging temperature dependence for the susceptibility above the Curie temperature with a critical exponent of two, in agreement with the experimental results.

## II. EXPERIMENT

Polycrystalline $(1\text{-}x)Pb(Mg_{1/3}Nb_{2/3})O_3$ – $xPbTiO_3$ samples with $x$ = 0, 0.25 and 0.35 were prepared by the ‘columbite method’,[5] which consisted of the synthesis of a columbite $MgNb_2O_6$ phase at 1100°C for 5 hours, followed by a final reaction with ($PbO$ + $TiO_2$) to form the perovskite phase. The mixture of ($MgNb_2O_6$ + $TiO_2$ + $PbO$) was calcinated at 940 °C for 4 hours, thoroughly re-grounded, and then sintered at 1230-1250 °C for 4 - 8 hours. X-ray diffraction confirmed the formation of a pure perovskite phase. The surfaces of the pellets were finely polished, ultrasonically cleaned and then painted with silver electrodes.

The complex dielectric permittivity was measured using a computer-controlled impedance analyzer (Solartron 1260) in conjunction with a dielectric interface (Solartron 1296) as a function of frequency at different temperatures under isothermal conditions. A weak (0.05-0.2 V/mm) ac field was applied. The data were taken upon cooling the sample starting from 720 K.

## III. EXPERIMENTAL RESULTS AND DISCUSSION

### A. Critical behavior of the imaginary susceptibility in PMNT75/25

Determining the universal susceptibility $\chi_U$ in relaxors does not appear to be a simple task because of the presence of other polarization mechanisms besides the universal polarization. In PMNT75/25, at $T > T_m$ the measured real part of permittivity $\varepsilon'$ is the sum of $\chi'_U$ and the

conventional relaxor permittivity ($\varepsilon'_R$). In addition, at extremely low frequencies and comparatively high temperatures, the susceptibility related to the polarization of the slowly mobile charge carriers also become significant.[2-4] To extract $\chi'_U$, one can use the characteristic dependences of the real and imaginary components on frequency ($f$), which are given by the fractional power-law relations[2-4]

$$\chi''_U(f) \propto f^{n-1}, \qquad (3)$$

$$\chi'_U(f) = \tan(n\pi/2)\chi''_U(f), \qquad (4)$$

where $n$ is the temperature-dependent parameter that is close to, but smaller than, unity. In PMNT75/25 the values of $n$ at some particular temperatures were determined from the frequency dependences of the imaginary part of the permittivity using Eq. (3) in those frequency intervals where $\chi''_U$ was the single contribution to the losses, and then the values of $\chi'_U(T,f)$ were calculated using Eq. (4) (see Refs. 2 and 3 for the details). But in the cases of PMNT65/35 and PMN, these frequency intervals are too narrow and $n$ (and therefore $\chi_U'$) values cannot be determined with a good precision (see below). Consequently, the validity of Eq. (1) cannot be experimentally verified. In this section we will show that this difficulty can be overcome. It turns out that the critical behavior can be detected not only for the real part of the universal relaxor susceptibility, but also for the imaginary part.

The real component follows the critical temperature dependence (1), and according to Eq. (4), varies in proportion to the imaginary one with the proportionality factor depending on $n$. Thus, if $n$ were constant (i.e. independent of temperature), $\chi''_U(T)$would also follow the critical dependence

$$\chi''_U = C''_\chi (T - T_0)^{-\gamma} \text{ at } T > T_0, \qquad (5)$$

with the same parameters $T_0$ and $\gamma$, but different $C''_\chi = C'_\chi \cot(n\pi/2)$. However, $n$ is not really a constant. The temperature dependence of $n$, determined in PMNT75/25 ceramics by fitting the experimental $\chi''_U(f)$ data to Eq.(3) (see Ref. 2 and 3 for the details), is shown in Fig. 1. Only at high enough temperatures the value of $n$ changes slowly with temperature and thus $\chi''_U(T)$ should approximately satisfy Eq. (5). To verify that, the experimental $\chi''_U(T)$ dependences of PMNT75/25 ceramics were fitted to Eq. (5) at several frequencies using the least squares technique. To avoid any significant influence of the inconstancy of $n$, the temperature intervals of $T > 400$ - $410$ K were chosen for fitting. The best-fit $\gamma$ values were found to be equal to 1.97-2.15 with the tendency to grow with decreasing frequency or temperatures at which the fitting was performed. This tendency can be explained by the temperature variation of $n$. The parameters $T_0$ and $C''_\chi$ are also dependent on frequency with the best-fit values varying from $T_0 = 367$ K and $C''_\chi = 9\times10^5$ K$^2$ at 0.1 Hz to $T_0 = 383$ K and $C''_\chi = 7\times10^4$ K$^2$ at 10 kHz. Fig. 2 illustrates the results of fitting for selected frequencies. One can see that in a wide temperature range, $\varepsilon''$ (= $\chi''_U$) follows well Eq. (5) with $\gamma \cong 2$, giving rise to the straight lines in the considered log-log presentation. The experimental relationship deviates from linear at high and low temperatures, first of all, as a result of the contribution from the aforementioned additional relaxation processes that exist besides the universal relaxor relaxation, leading to $\varepsilon'' \neq \chi''_U$. At high temperatures, the contribution that increases with increasing temperature but decreases with increasing frequency, results from the low-frequency dispersion (LFD) which is probably related to the motion of slowly mobile charge carriers.[2,3] The low-temperature deviation, when $\varepsilon''$ noticeably exceeds the $\chi''_U(T)$ trend predicted by Eq. (5), is the result of conventional relaxor dispersion (CRD). At temperatures close to $T_0$, a round-off of the data is observed and the values of $\varepsilon''$ become smaller than those predicted by Eq. (5), which means a violation of the critical behavior for $\chi''_U$, in consistence with the analogous violation of Eq. (1) previously observed in Ref. 4.

Therefore, in this section we have demonstrated, using the exprimental data of PMNT75/25, that if the real part $\chi'_U(T)$ diverges at $T_0$ according to Eq. (1), the corresponding imaginary part $\chi''_U(T)$ will also diverge at the same $T_0$ with the same value of critical exponent $\gamma$, provided the value of $n$ is temperature-independent.

### B. Universal relaxor dispersion and critical behavior in PMNT65/35

In the (1-$x$)PMN- $x$PT solid solutions, the morphotropic phase boundary is known to occur at $x_c \cong 0.3$,[6,7] which means that below the temperature of FE phase transition, the crystal structure is rombohedral at $x < x_c$, but tetragonal at $x > x_c$. One can expect that the dielectric behavior would be different in these two cases.

The temperature dependences of the reciprocal dielectric permittivity of PMNT65/35 ceramics (i.e. at $x > x_c$) in the temperature range below and slightly above $T_m$ are shown in Fig. 3. Unlike the samples with $x$=0.25 ($x < x_c$), where $\varepsilon(T)$ in the low-temperature FE (rhombohedral) phase can be described by the Curie-Weiss law (2),[4] this law is not valid in the case of PMNT65/35. The temperature variation of $1/\varepsilon'$ suggests a first-order (although slightly diffused) ferroelectric phase transition at approximately 450 K. This observation is in consistence with the X-ray data showing the abrupt change of unit-cell parameters at the Curie point.[6]

The frequency dependences of the imaginary part of the permittivity at temperatures around and above $T_m$ are shown in Fig. 4. Solid lines represent the fit to Eq. (3) performed in the frequency intervals where the losses are determined by the universal relaxor dispersion (URD) only (and thus $\chi''_U$ equals to the measured value of $\varepsilon''$). Qualitatively the same peculiarities as previously observed in PMNT75/25 ceramics[2-4] are found, namely, the validity of the relation (3) at certain frequency intervals (where $\varepsilon''= \chi''_U$), and the violation of Eq. (3) at low frequencies due to the LFD contribution, as well as at the temperatures close to $T_m$=450 K due to the CRD. The CRD appears on the high-frequency side of the spectrum and moves to lower frequencies with decreasing temperature. The distinction between PMNT65/35 and PMNT75/25 is that, the former has a much less pronounced CRD. The best-fit values of $n$ are found to vary between 0.98 and 0.94 with decreasing temperature. These values are close to those found for PMNT75/25 (see Fig. 1). Similar to PMNT75/25, not only the imaginary but also the real part of the permittivity in PMNT65/35 follows the fractional power law in the temperature-frequency region where URD is a single relaxation process (not shown). In this region the values of $\chi'_U$ can be extracted (as described in the previous section). However, because this region is comparatively narrow, the precision in the values of $n$ (and thus $\chi'_U$ ) so determined turns out to be quite low and the validity of the relation (1) cannot be reliably verified. Nevertheless, as demonstrated in the previous section, if $n(T)$ changes slowly and Eq. (1) holds, Eq. (5) should also hold.

Fig. 5 shows the temperature dependences of the imaginary part of permittivity in PMNT65/35 ceramics obtained by fitting the experimental data to Eq. (5) in the same manner as it was done in Sec. III A. It can be seen that Eq. (5) is satisfied in a wide temperature range with the same features as in PMNT75/25, namely, (i) the best-fit values of $\gamma$ are close to 2 (between 2.0 and 2.11), (ii) $C_\chi''$ decreases from $2\times10^5$ K$^2$ at 100 Hz to $6\times10^4$ K$^2$ at 100 kHz, and (iii) $T_0$ increases from 428 to 439 K in this frequency interval. The high-temperature deviation from Eq. (5) related to LFD is also evidenced. At low frequencies ($f <$ 100 Hz) LFD begins at temperatures close to $T_0$. As a result, the losses are determined by the URD only in the very narrow temperature intervals and the validity of Eq. (5) cannot be checked reliably.

It is shown in this section that the URD and the critical quadratic behavior of the universal relaxor susceptibility as a function of temperature are observed in (1-$x$)Pb($Mg_{1/3}Nb_{2/3}$)$O_3$ – $x$$PbTiO_3$ ceramics with $x$ (=0.35) > $x_c$, where a first-order phase transition to the tetragonal ferroelectric phase takes place and the Curie-Weiss law is not observed at $T < T_m$.

### C. Universal relaxor dispersion and critical behavior in PMN

Contrary to the solid solutions studied above, there is no transition at all to a ‘normal’ ferroelectric phase in pure PMN. A macroscopically isotropic relaxor state exists below the diffused phase transition temperature.[1] At $T > T_m \cong 270$ K the measured complex dielectric spectra suggest the presence of several overlapping relaxation processes. The example for room temperature is shown in Fig. 6. At extremely low frequencies, the relation $\varepsilon''(f) \propto 1/f$ is observed, indicating the contribution of dc conductivity. The relaxation process with a characteristic (peak) frequency $f_p$ of about 0.1 Hz is also detected. The dispersion in the same temperature-frequency range was previously reported for PMN crystal.[8] The fractional power-law frequency relation $\varepsilon''(f) \propto f^{n-1}$ with $n = 0.52$ is found at $f_p < f < 100$ Hz (see the dashed line in Fig. 6). Satisfaction to this law is a usual behavior for the losses in dielectrics above the peak frequency.[9] At frequencies higher than about 100 Hz, the $\varepsilon''(f)$ pattern changes: the fractional power law still remains valid, but with a different exponent parameter, $n = 0.74$ (represented by the solid line). This suggests the existence of another relaxation process. For the reasons discussed below we believe that it is the same URD as observed in the (1-$x$) PMN - $x$PT samples.

Unlike in PMNT75/25 and PMNT65/35 where the range of the CRD shifts to lower frequencies with decreasing temperature,[3] in PMN ceramics, it consists of two (a low-frequency and a high-frequency) branches, and the high-frequency one is practically temperature independent.[10] In Fig. 6, we observe the tail of the low-frequency branch at $f > 10^4$ Hz, which has hidden the URD at those frequencies.

All the three relaxation processes mentioned above can be clearly identified in the complex plane representation of $\varepsilon''(\varepsilon')$. Fig. 7 shows only the high-frequency part of the plot in order to make the URD visible on an enlarged scale. In accordance with Eq. (4), the $\varepsilon''(\varepsilon')$ relation for the URD is linear (solid line in Fig. 7).

Thus, at room temperature the URD provides the main contribution to the total $\varepsilon''$ (i.e. $\chi''_U \approx \varepsilon''$) only in the frequency interval of about $10^2 - 10^4$ Hz. With increasing temperature, the lower –frequency limit of this interval increases due to an overlap of the accompanying dispersion. Consequently, even if $\chi''_U$ follows Eq. (5), the analogous critical behavior for the measured $\varepsilon''$ can be expected only in a narrow frequency range close to $10^4$ Hz.

Fig. 8 shows the $\log_{10}\varepsilon''$ vs. $\log_{10}(T - T_0)$ plot for the frequency $10^4$ Hz, which is obtained by fitting the experimental data of PMN to Eq.(5) with $\chi''_U = \varepsilon''$. The solid line illustrates the quality of the fit. The best-fit parameters, $\gamma$, $T_0$ and $C''_\chi$, are found to equal 2.06, 271 K and $3.2\times10^4$ K, respectively. Thus the experimental data within a wide temperature range from 280 K to 420 K follow the critical relation (5) with $\gamma \cong 2$, i.e. show the behavior typical of the universal relaxor permittivity previously found in PMNT. This fact confirms that the losses at $f \sim 10^4$ can really be attributed to the URD mechanism. At lower or higher frequencies the quadratic critical temperature dependence of $\varepsilon''$ is violated because not only the universal relaxor polarization, but also the polarizations of the other types, contribute to $\varepsilon''$.

In summary, we have shown in this section that, in pure PMN relaxor also, the universal relaxor polarization exists and follows the critical quadratic temperature behavior. Thus, the phase transition to a normal ferroelectric phase is not a necessary condition for the URD to occur and to exhibit the critical behavior in relaxor ferroelectrics.

## IV. MICROSCOPIC MODELING

The main experimental finding of the present work is that the diffused $\varepsilon(T)$ maximum in PMN, as well as PMNT, relaxors contains at least two contributions, one of which, called the universal relaxor susceptibility, obeys the critical relations (1) and (5) with $\gamma \cong 2$ at high enough temperatures. In this section, we intend to establish a microscopic model to explain such a behavior. Let's recall that the same value of critical exponent $\gamma$, describing the temperature dependence of the susceptibility above the Curie point, has been deduced from an approximation to the Ising model, known as the spherical model.[11,12] The Hamiltonian of this model has the form identical to the well-known Ising Hamiltonian[12]

$$H = -J\sum_{ij} S_i S_j \,, \qquad (6)$$

where the parameter $J$ describes the interaction between pseudospins $S_i$. But unlike in the Ising model, where $S$ can be equal to +1 or -1 only, the magnitude of $S$ in the spherical model is no longer a constant of the motion. It can take different values, subject only to the constraint,

$$\sum_i S_i^2 = N \,, \qquad (7)$$

where $N$ is the total number of pseudospins in the system.

The above-mentioned similarity in the critical behavior suggests that the universal relaxor polarization may be described by the standard spherical model. We now attempt to prove this based on the crystal chemistry feature and microscopic dipole interactions of the relaxors. Note that the so-called spherical random bond – random field (RBRF) model was recently proposed to explain the NMR data and the non-linearity of the total dielectric susceptibility in relaxors.[13,14] The linear susceptibility measured in the present work, if analyzed by the RBRF model, however, does not obey Eqs. (1) and (5) (this will be discussed in more detail below).

It was established[1,15] that in relaxor materials, polar regions of nanometer size (polar nanoregions, PNRs) begin to appear and grow when temperature decreases below a certain Burns temperature $T_B$. These regions are embedded into non-polar surroundings and randomly oriented. The experiments show that in PMN, $T_B$ exists at about 600-630 K, i.e. at temperatures much higher than $T_m$. The published data on the values of $T_B$ in (1-$x$)PMN- $x$PT are scarce, but it seems that $T_B$ is practically independent of $x$, at least up to $x = 0.4$. It was determined from the temperature dependences of the lattice parameters, which deviate from a straight line at temperatures below about 600 K.[6] We have suggested before[3] that the universal relaxor susceptibility originates from the change of PNR polarization. The mechanisms underlying such a picture are being analyzed comprehensively in this section.

Although the existence of PNRs in relaxors is well documented, their structure and origin are still the subjects of controversial speculations. In the early works, it was suggested that due to the compositional disorder in the arrangement of different ions (e.g. $Mg^{2+}$ and $Nb^{5+}$ in PMN) on the equivalent crystallographic sites, the ferroelectric Curie temperature in the relaxors is subject to strong frozen spatial fluctuations and the PNRs are just the regions of higher Curie temperature.[16] Upon cooling, the local ferroelectric phase transitions occur in these regions first, whereas the other parts of the crystal remain in the paraelectric phase. This approach implies that the PNRs in relaxors have the structure analogous to the structure of a FE phase, i.e. all the unit cells inside the polar region are spontaneously polarized in one and the same direction. Accordingly, the magnitude of PNR dipole moment scales with the PNR size (i.e. the number of the polar unit cells therein). Similar structure of PNRs, i.e. a ferroelectric-type ordering of the unit cell dipole moments, was postulated in other models developed for the description of the relaxor state.

The key distinction of our approach from the previous ones is that we assume the possibility for the dipole moment of a unit cell *inside* a PNR to take different directions with respect to the dipole moment of the PNR itself. The following arguments can be adduced in support of the assumption. It is known that the degree of compositional disorder can influence greatly the type of the dipole ordering. Complex perovskites in which different cations are arranged in order on the equivalent crystallographic sites, thus creating a superstructure, usually exhibit an antiferroelectric (AFE) phase below the Curie temperature. In the compositionally disordered complex perovskites, the AFE phase has never been found. Instead, the FE or relaxor FE properties have been observed. Furthermore, for those crystals, the degree of compositional disorder of which can be changed by appropriate high-temperature treatments, the same sample can possess an AFE or an FE dipole ordering in the compositionally ordered or disordered state, respectively.[17]

The direction of the spontaneous displacement of a certain ion from a high-symmetry position in ferro- or antiferroelectrics below the Curie temperature (or inside the polar region of a relaxor below $T_B$) is determined by the balance of short-range and long-range (dipole-dipole) forces acting on that ion. These forces, in turn, are determined by the composition of unit cells surrounding the ion in consideration (primarily the nearest neighboring cells). For example, when the $A(B'B'')O_3$ complex perovskite is compositionally ordered (with B' and B" alternating along cubic <100> directions), any cell (which is referred to the primitive perovskite cell) containing B' cations has as its nearest neighbors only the cells with B" cations and *vice versa*. Such an ionic configuration favors the occurrence of an AFE dipole order, as discussed above. In the compositionally disordered state, the B cations of different types are located in certain neighboring unit cells causing the appearance of an FE ordering. But even in the case of a full disorder, the cells with one certain type of ions (say B') having as its nearest neighbors only the cells with the other type of ions (B") still exist (although in a rather small concentration). Some other unit cells may border with one or more cells containing the same type of B ions (the formulas for calculating the concentration of such cells are reported e.g. in Ref. 18). One can expect that at those points inside a PNR, which contain the unit cells surrounded by a majority of the cells of the different type, in the course of dipole ordering, the cations shift in a direction opposite to that of the majority surrounding cations, i.e. as in the case of the AFE structure (or it is possible that the dipole moments in the neighboring unit cells are non-collinear). On the contrary, the cells inside the PNR, which are enclosed by a large number of the cells with the same B cations, favor an FE order. Due to these FE or AFE interactions, the dipole moments of the above-mentioned cells are locally fixed relative to the directions of the dipole moments in the neighboring cells and an external field can switch them simultaneously only. These cells are called "fixed" cells. On the other hand, the cells with an intermediate number of identical bordering cells may show two (or more) different directions of dipole moment with the same (or almost the same) energy. The moments of these "free" cells can flip from one potential well to another while the moments of neighboring cells remain unchanged. The planar presentation of the above-described compositional and dipole configurations is shown in Fig. 9.

There is of course no definite boundary between the “fixed” and “free” cells; the cells with intermediate properties can actually exist, but for the sake of simplicity, we assume that all the cells of a PNR can be divided into these two groups. For the same reason, we will consider below the one-dimensional variant of the dipole ordering.

The above-drawn picture implies that a glassy state with nonzero spontaneous polarization exists inside the polar region. But an important feature of this state is that it exists inside the very small volumes of PNRs, surrounded by an adjacent paraelectric phase.

The idea of the short-range AFE ordering in the structure of relaxors has found its experimental basis in recent studies of PMN by selected-area electron diffraction and high-resolution electron microscopy,[19] and in x-ray diffuse scattering studies of the another relaxor $Pb(Sc_{1/2}Nb_{1/2})O_3$,[20] in which the existence of antiphase shifts of ions of neighboring unit cells have been evidenced.

Based on the structural model proposed, we can now provide a description of the diffuse phase transition in relaxors. Upon cooling below $T_B$, the nanometer-sized polar regions begin to appear in the paraelectric matrix. The total dipole moment of each PNR ($P_i$ ) results from the moments of both the "fixed" and "free" unit cells. The "fixed" cells form a "rigid" framework (cluster) so that their moments can change the directions only simultaneously together with $P_i$ (as in the case of normal FE domains). Each free moment inside the polar region can fluctuate between the different equilibrium positions with the same (or almost the same) energy, while the direction of $P_i$ remains unchanged.

Consider a crystal containing $N$ polar nanoregions with the dipole moments $P_i$, $i = 1, 2, \ldots, N$. Assuming that the magnitudes of the dipole moment ($p$) of both the free and fixed cells are the same, the moment of $i$th polar region at a certain time can be represented as

$$P_i = (n_i - m_i + 2k_i - l_i)p\,, \tag{8}$$

where $n_i$ and $m_i$ are the numbers of fixed cells in the $i$th PNR with the dipole moments parallel and antiparallel to $P_i$, respectively, $l_i$ is the total number of free cells in the $i$th PNR and $k_i$ is the fraction of these free sells, which are oriented parallel to $P_i$ (so with $k_i \le l_i$).

Note that in the $Pb(Mg_{1/3}Nb_{2/3})O_3$–type complex perovskite relaxors, different unit cells contain differently charged cations ($Nb^{5+}$ and $Mg^{2+}$) and hence have different values of $p$. It was argued[14] that the role of Mg-containing cells is negligible because of the ferroelectrically inactive nature of $Mg^{2+}$. Therefore, the numbers $n_i$ , $m_i$, $l_i$ and $k_i$ refer to the Nb-containing cells only.

Due to the easy flipping of the free cells, the magnitude of $P_i$ can fluctuate around the equilibrium value of $|P_i| = (n_i - m_i)|p|$, with

the upper and lower limited values defined by the relations $|P_i| \le (n_i - m_i + l_i)|p|$; $|P_i| \ge (n_i - m_i - l_i)|p|$ [or $|P_i| \ge 0$, if $l_i \ge (n_i - m_i)$], respectively.

The moment of the PNR in a small electric field $E$ (which may be an external field applied to the crystal or an internal field arising from the interactions with the other PNRs) can be alternatively written as

$$P_i = (n_i - m_i)p + \alpha l_i E\,, \quad (9)$$

where $\alpha$ is the average polarisability of the free cells, and from Eqs. (8) and (9) we have,

$$2k_i - l_i = l_i \alpha E / p\,. \quad (10)$$

As discussed above, the values of $|P_i|$ can vary in a wide range beginning from zero. To show that the dipole moments of PNRs can be considered as an order parameter field of the spherical model, one needs to find the variable that can change while remaining proportional to $P_i$ and satisfy Eq. (7). Such a variable can be written as

$$S_i = sP_i / p\,, \quad (11)$$

with $s$ given by

$$s = \left[\left\langle (n_i - m_i + 2k_i - l_i)^2 \right\rangle\right]^{-1/2} = \left[N \Big/ \sum_i (n_i - m_i + 2k_i - l_i)^2\right]^{1/2},$$

where <…> denotes the averaging over all PNR's. Taking into account Eq. (10), we can find for the crystal under a small uniform field that

$$s^{-2} = \left\langle (n_i - m_i)^2 \right\rangle + \left\langle 2(n_i - m_i) l_i \alpha E / p \right\rangle. \quad (12)$$

Above the temperature of the spherical model phase transition, the directions of $P_i$ are randomly distributed with respect to $E$ and thus the second average term in Eq. (12) equals to zero, because the positive or negative values of $p$ should be taken when averaging, depending on the direction of the corresponding polar region moment. Thus, $s$ is a constant and $S_i$ is proportional to $P_i$. The validity of (7) can also be easily verified.

One might think at first glance that the spherical model is not physically plausible, because the constraint (7) allows the magnitude of one pseudospin to be very large (up to $N^{1/2}$) and that of other pseudospins to be very small and even zero. But, by analyzing Hamiltonian (6), one can show that states of this kind have comparatively high energy and thus do not contribute significantly into the partition function. Therefore, the system tends to stay in those states with close values of $S_i$, which mimic a real system of PNRs.

Thus, we have demonstrated that the PNRs in relaxor ferroelectrics can indeed be considered as the pseudospin of the standard spherical model. In contrast to the usual Ising model which cannot be exactly solved for the case of three-dimensional lattice, the partition function of the spherical model has been calculated, the exact equation of states has been derived and the following relation for the susceptibility in the high-temperature phase ($T > T_0$) has been obtained:[12]

$$\chi = (2Jw)^{-1}, \quad (13)$$

where $w$ is a function of temperature which vanishes when $T \to T_0$ as $w \propto (T - T_0)^2$. The present experimental findings show that the temperature dependence of the universal susceptibility $\chi_U$ exhibits the same behavior. This, in turn, confirms that the interacting PNRs are the origin of the universal relaxor susceptibility observed in the high-temperature ($T > T_m$) phase of the relaxors.

When temperature decreases, the number and the size of PNRs are known to grow[1] and consequently the changes of $J$ and $s$ can be expected. The temperature dependence of susceptibility (13), the proportionality between $P_i$ and $S_i$, and other conditions of the spherical model become disturbed in this case. This can explain why Eqs. (1) and (5) describing the universal susceptibility become violated when the temperature decreases approaching the phase transition (see Figs. 2, 5 and 8) and why the phase with a long-range ferroelectric order predicted by the spherical model, is not actually developed in relaxors, such as pure PMN.

In the proposed model, the dipole moments of the free cells (which occupy the most part of the PNRs) can fluctuate between the different states separated by comparatively small potential barriers, or even tunnel between them. The height of the barriers may vary in a wide range from one cell to another as well as in the course of time. This implies the existence of a wide relaxation spectrum, which results in the universal relaxor dispersion. To describe this dispersion, a dynamic theory should be developed in the future.

As mentioned above, the RBRF model of relaxors has recently been proposed.[13,14] The RBRF is an extension of the standard spherical model, where the randomness of the interaction strengths $J$ between different pseudospins is taken into account and one more term in the Hamiltonian is added, which describes the random fields caused by the disorder in the ionic structure. This model was applied to the PNR dipoles that reorient as a whole, and verified by comparing with the total (but not the universal) susceptibility. It is shown in Refs. 2,4 and in this work that the conventional relaxor susceptibility is the dominant part of the total one. In the present study, we consider another polarization mechanism giving rise to the universal relaxation. This mechanism has nothing to do with the large potential barriers for $P_i$ flipping considered by the RFRB model. It rather deals with an extremely soft subsystem of free cells with a wide spectrum of low energy potential barriers. This justifies using the standard spherical model to describe the temperature evolution of the universal susceptibility. Thus, our new approach and the RFRB model do not contradict each other. Instead, they merely describe the different parts of the dielectric response in relaxors.

## VI. CONCLUSIONS

In this work we have performed the dielectric spectroscopic studies to determine whether the universal relaxor dispersion recently discovered in PMNT75/25 ceramics near and above $T_m$, also exists in the PMN-PT solid solutions of other compositions and in pure PMN. We have first demonstrated that, if the real part of the universal relaxor susceptibility $\chi'_U(T)$ diverges at $T_0$ according to Eq. (1), the corresponding imaginary part $\chi''_U(T)$ will diverge also at the same $T_0$ with the same value of critical exponent $\gamma$, provided the value of $n$ is temperature-independent. This conclusion has been drawn from the experimental verification of Eq. (5) in PMNT75/25 at the temperature range where the condition of $n$ constancy is approximately met (i. e. well above $T_0$). In the PMNT65/35 and PMN ceramics, the universal relaxation process with losses described by Eq. (3) has been found. Due to the overlap of the URD with the other types of relaxation processes in PMN and PMNT65/35, the values of the real part of the universal susceptibility $\chi'_U$ (which is a relatively small part of the total permittivity) cannot be accurately determined in a temperature range wide enough to verify the validity of Eq. (1). Nevertheless, the imaginary part $\chi''_U$ in these compounds was found to follow Eq. (5) with $\gamma \cong 2$, at least for the frequency ranges where $\chi''_U$ is the single contribution to the losses. This implies that Eq. (1) with $\gamma \cong 2$ should also hold for PMN65/35 and PMN.

These results indicate that the critical behavior of the universal relaxor susceptibility, as described by Eqs. (1) and (5), can be observed in different relaxor ferroelectrics, ranging from the prototypical relaxor PMN to the materials in which the relaxor state spontaneously transforms to a normal ferroelectric state via a second-order (PMNT75/25) or first-order (PMNT65/35) phase transition. It is expected to be one of the common properties of the relaxor ferroelectrics.

Finally, to describe the microscopic origin of the universal relaxor polarization, we have suggested the model relating this polarization to the microscopic dipole moments of the mutually interacting PNRs, which exist in relaxors in a wide temperature range starting from the temperatures far above $T_m$. The model assumes that each PNR consists of differently polarized unit cells, including (i) the cells with spontaneous dipole moment along the same direction as the total moment $P_i$ of the PNR, (ii) the cells with the moments fixed in the opposite direction, and (iii) the cells that are free to choose different directions with respect to $P_i$. Because of such a local structure of PNR, the magnitude of $P_i$ can vary within a wide range of values. This feature has allowed us to describe the system of interacting PNRs using the standard spherical model, which predicts a quadratic divergence of susceptibility above the Curie temperature, in agreement with the experimentally observed temperature behavior of the universal relaxor susceptibility. This result has revealed that the universal relaxor polarization is directly related to the microscopic chemical and polar structures inherent in relaxors, and hence is one of the main features of the relaxor ferroelectrics.

## ACKNOWLEDGMENTS

The authors would like to thank Dr. S. A. Prosandeev for many stimulating discussions and fruitful suggestions. This work was supported by the U.S. Office of Naval Research (Grant No. N00014-99-1-0738).

## Figure Captions:

FIG. 1. Temperature dependence of the exponent *n* of the universal relaxation in the PMNT75/25 ceramics.

FIG. 2. Fitting of the imaginary component of the permittivity ($\varepsilon'' = \chi_U''$) of the PMNT75/25 ceramics to Eq.(5) at selected frequencies. Experimental values of $\varepsilon''$ (points) are plotted against $\log(T - T_0)$ with the best-fit critical temperatures $T_0$=382.5, 375, 374 and 367 K for the frequencies of $10^4$, $10^2$, 10 and 0.1 Hz, respectively.

FIG. 3. Reciprocal of the real part of the permittivity of the PMNT65/35 ceramics as a function of temperature upon cooling for selected frequencies.

FIG. 4. Frequency dependences of the imaginary part of the dielectric permittivity of the PMNT65/35 ceramics at selected temperatures. Solid lines are the fits to Eq. (3) in the frequency intervals, where $\varepsilon'' = \chi_U''$.

FIG. 5. Fitting of the imaginary component of the permittivity ($\varepsilon'' = \chi_U''$) of the PMNT65/35 ceramics to Eq.(5) at selected frequencies. Experimental values of $\varepsilon''$ (points) are plotted against temperature with the best-fit critical temperatures $T_0$=439, 436, and 428 K for the frequencies of $10^5$, $10^4$, and $10^2$ Hz, respectively.

FIG. 6. Frequency dependence of the imaginary part of the dielectric permittivity of the PMN ceramics at $T$ = 295 K. Solid and dashed lines represent the fractional power law dependences $\varepsilon'' \propto f^{n-1}$ with $n$ equal to 0.74 and 0.52, respectively.

FIG. 7. Imaginary vs real part of permittivity of PMN ceramics at $T$ = 295 K. Solid line is the fit to Eq. (4) in the frequency intervals where $\varepsilon'' \approx \chi_U''$. Frequencies are indicated for selected points.

FIG. 8. Fitting of the imaginary component of permittivity ($\varepsilon'' \approx \chi_U''$) of PMN ceramics to Eq.(5) at $f$ = 10 kHz. Experimental values of $\varepsilon''$ (points) are plotted against temperature with the best-fit critical temperature $T_0$=271 K.

FIG. 9. Schematic illustration of the proposed model of the interacting PNRs in the compositionally disordered A(B'B")$O_3$ perovskite structure. A and O ions are not shown. Small arrows represent spontaneous dipole moments of "fixed" unit cells (one-end arrows) or "free" unit sells (double-end arrows). Large arrows indicate the dipole moments $P$ of individual PNRs. It is assumed in this example that the cell is ferroelectrically ordered (i.e. the direction of its moment is the same as the direction of $P_i$) if it has three or four B cations of the same type in the neighboring cells, and that the direction of the moment of a cell is opposite to $P_i$ if it has no B cations of the same type in the neighboring cells. "Free" cells have one or two B cations of the same type in the neighboring cells.

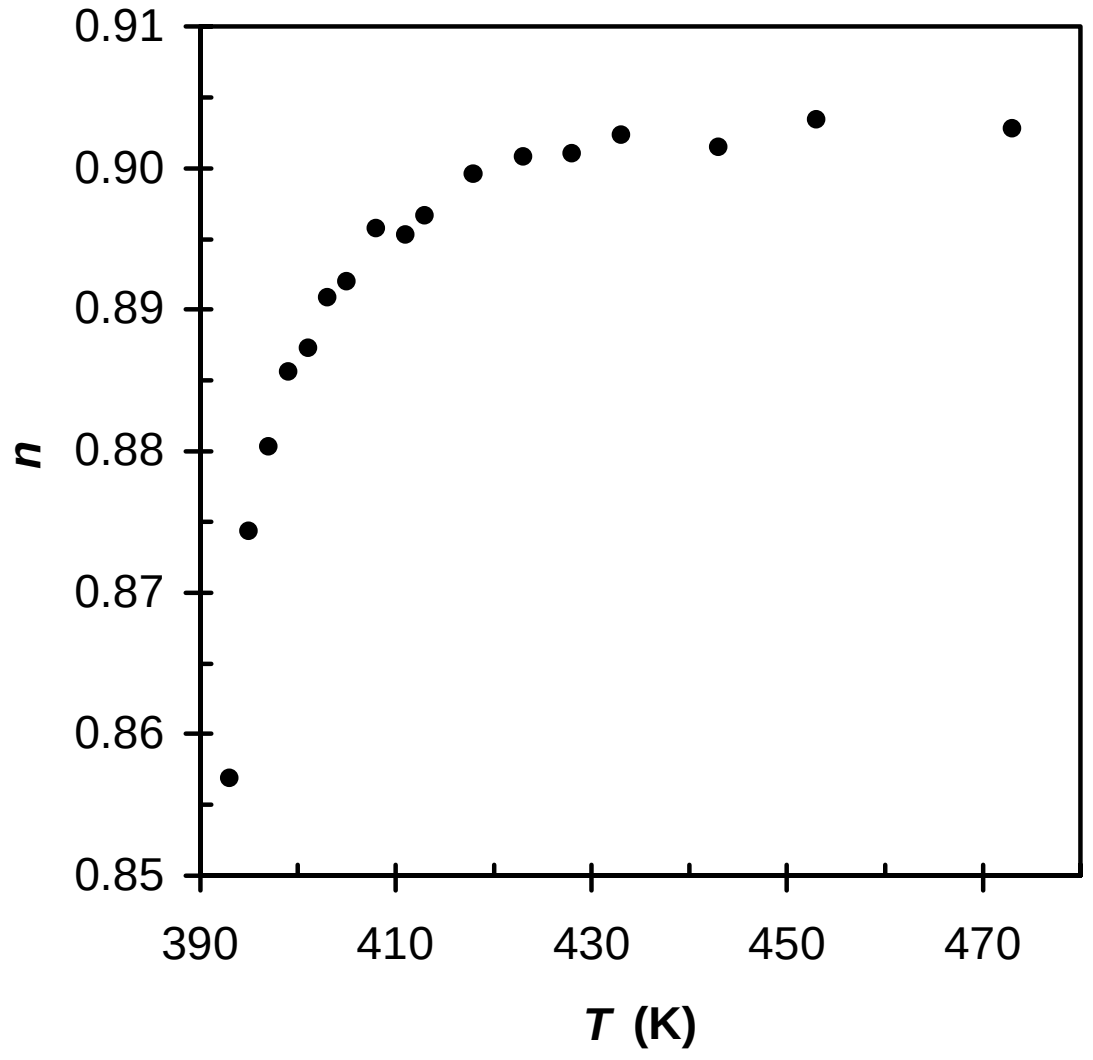

Fig. 1 (Bokov & Ye)

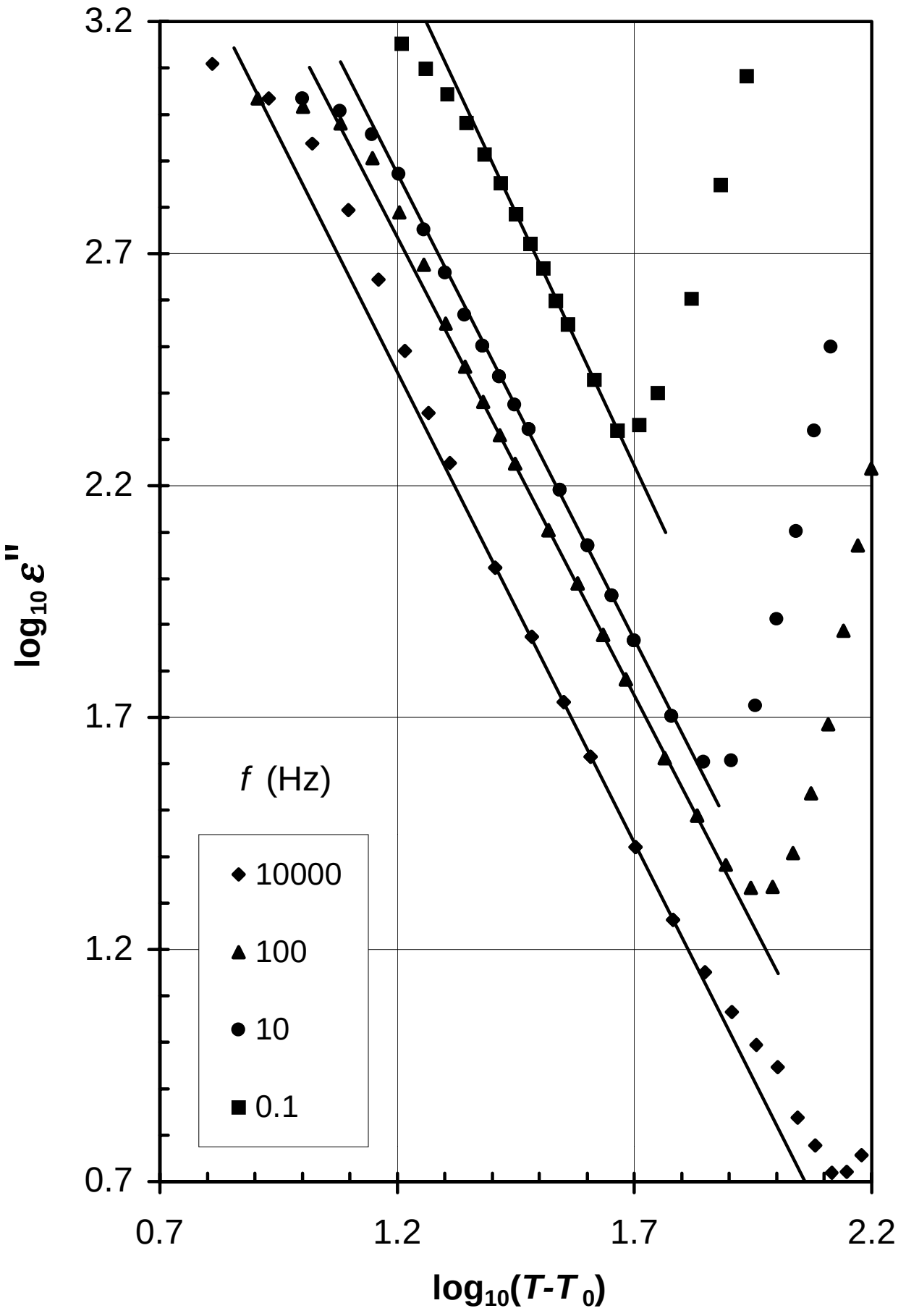

Fig. 2 (Bokov & Ye)

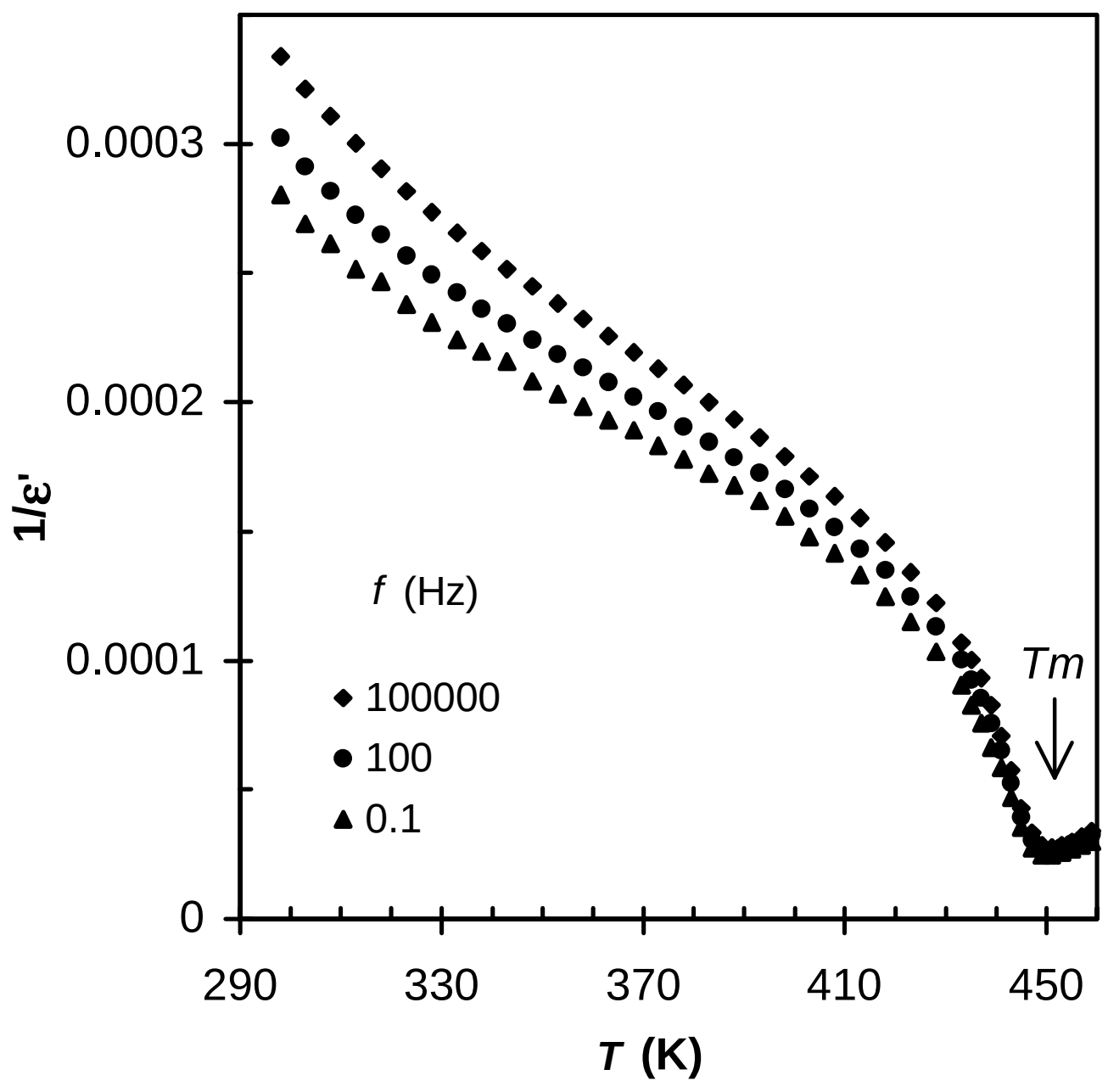

Fig. 3 (Bokov & Ye)

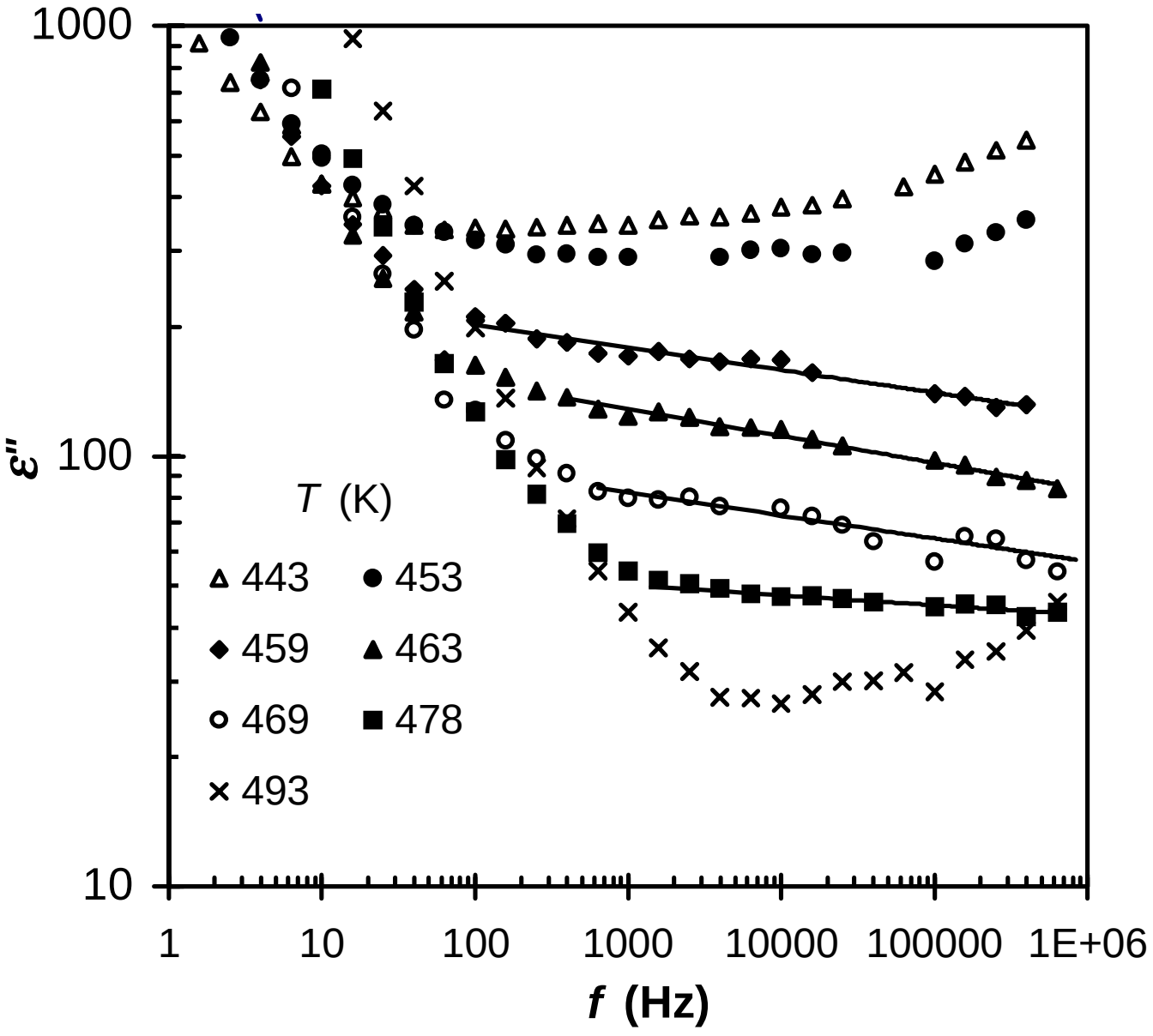

Fig. 4 (Bokov & Ye)

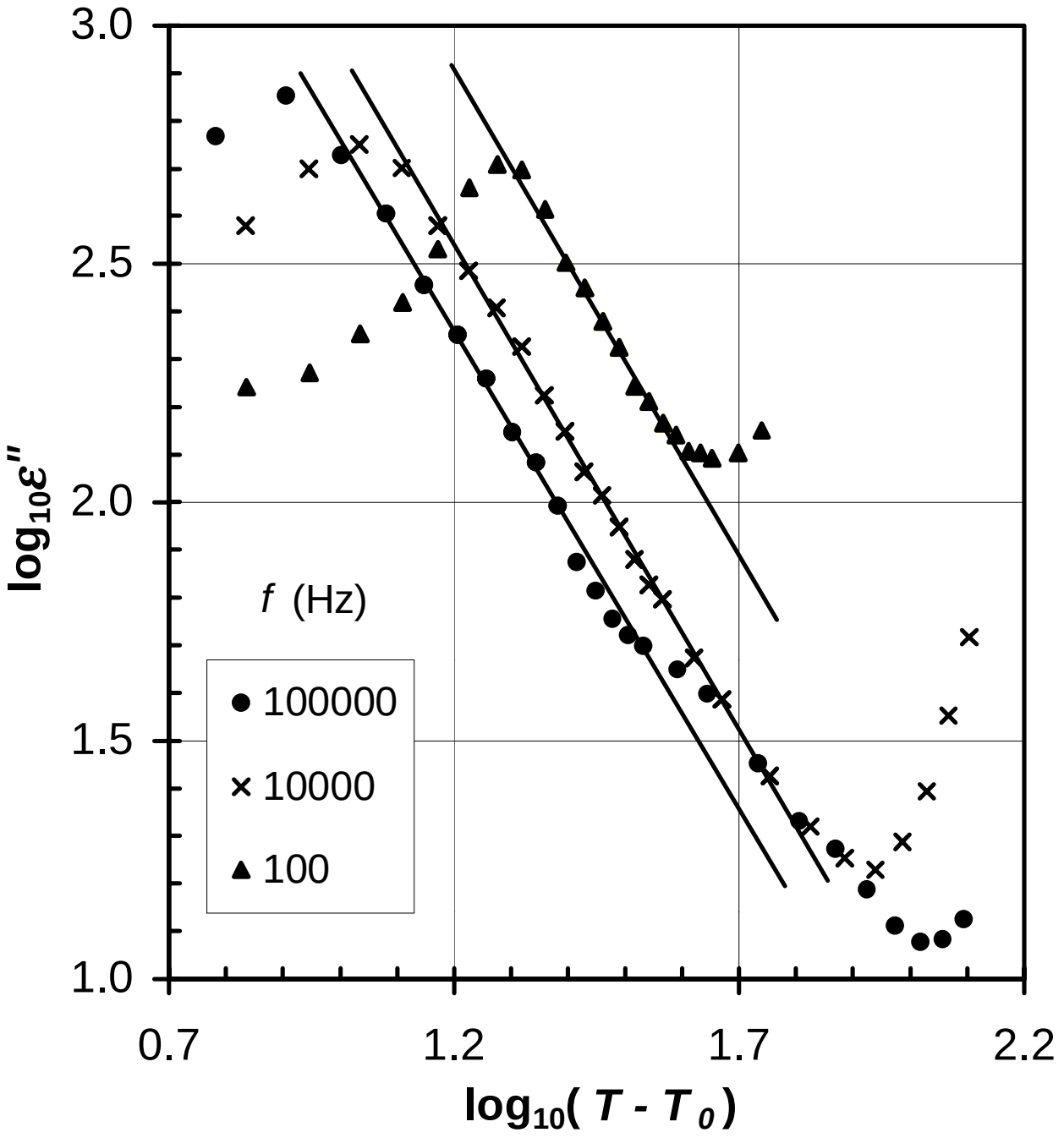

Fig. 5 (Bokov & Ye)

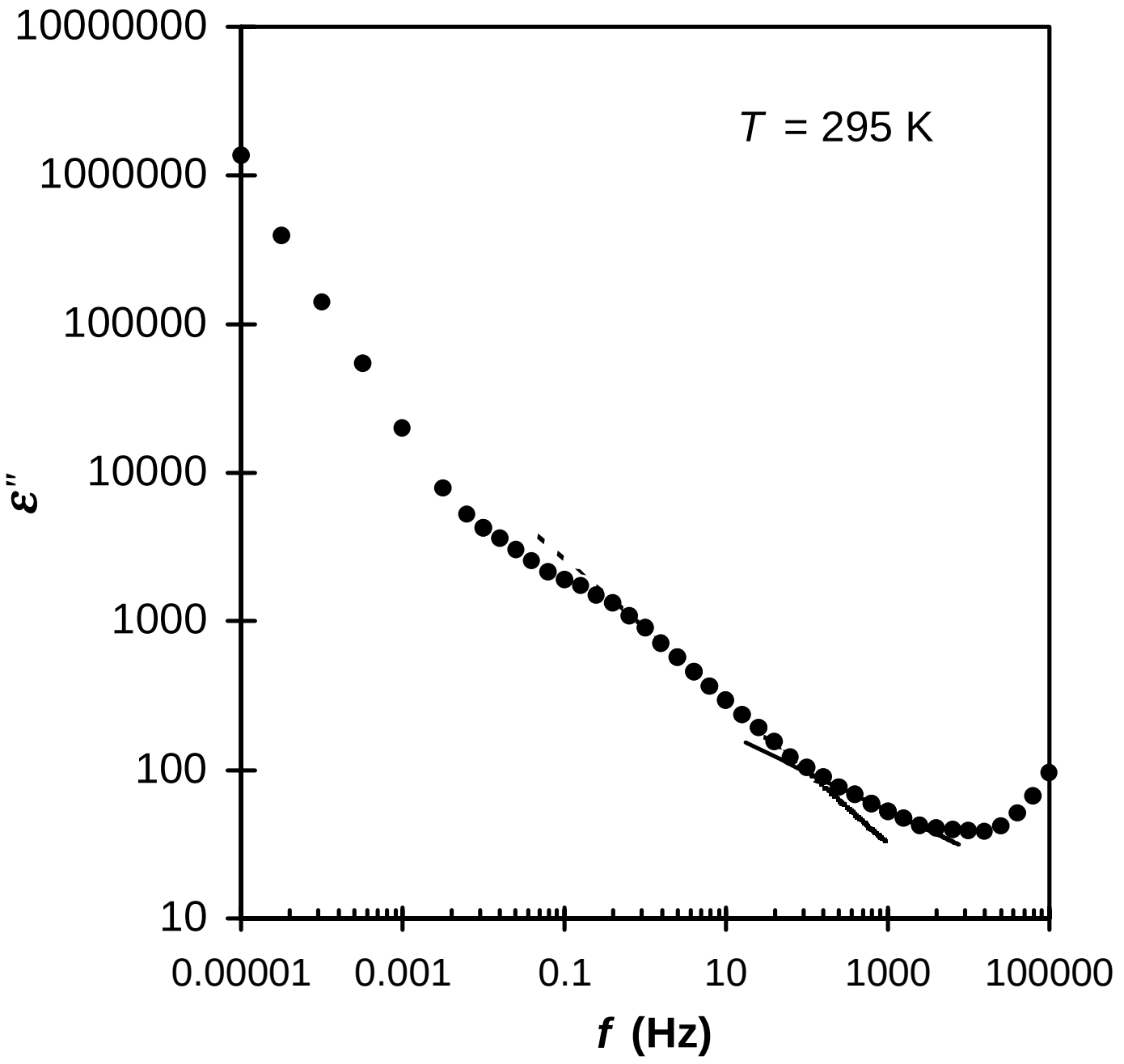

Fig. 6 (Bokov & Ye)

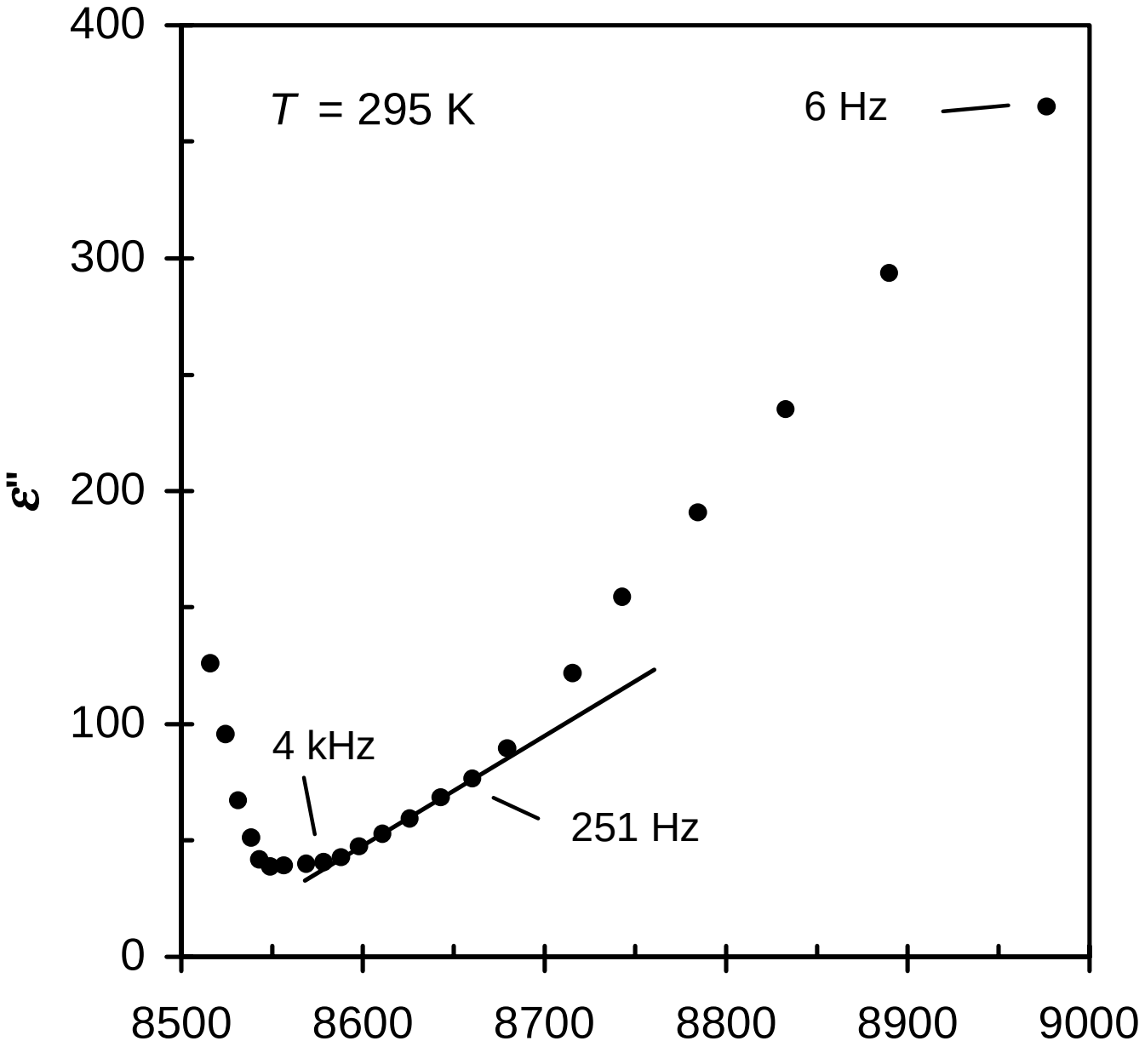

Fig. 7 (Bokov & Ye)

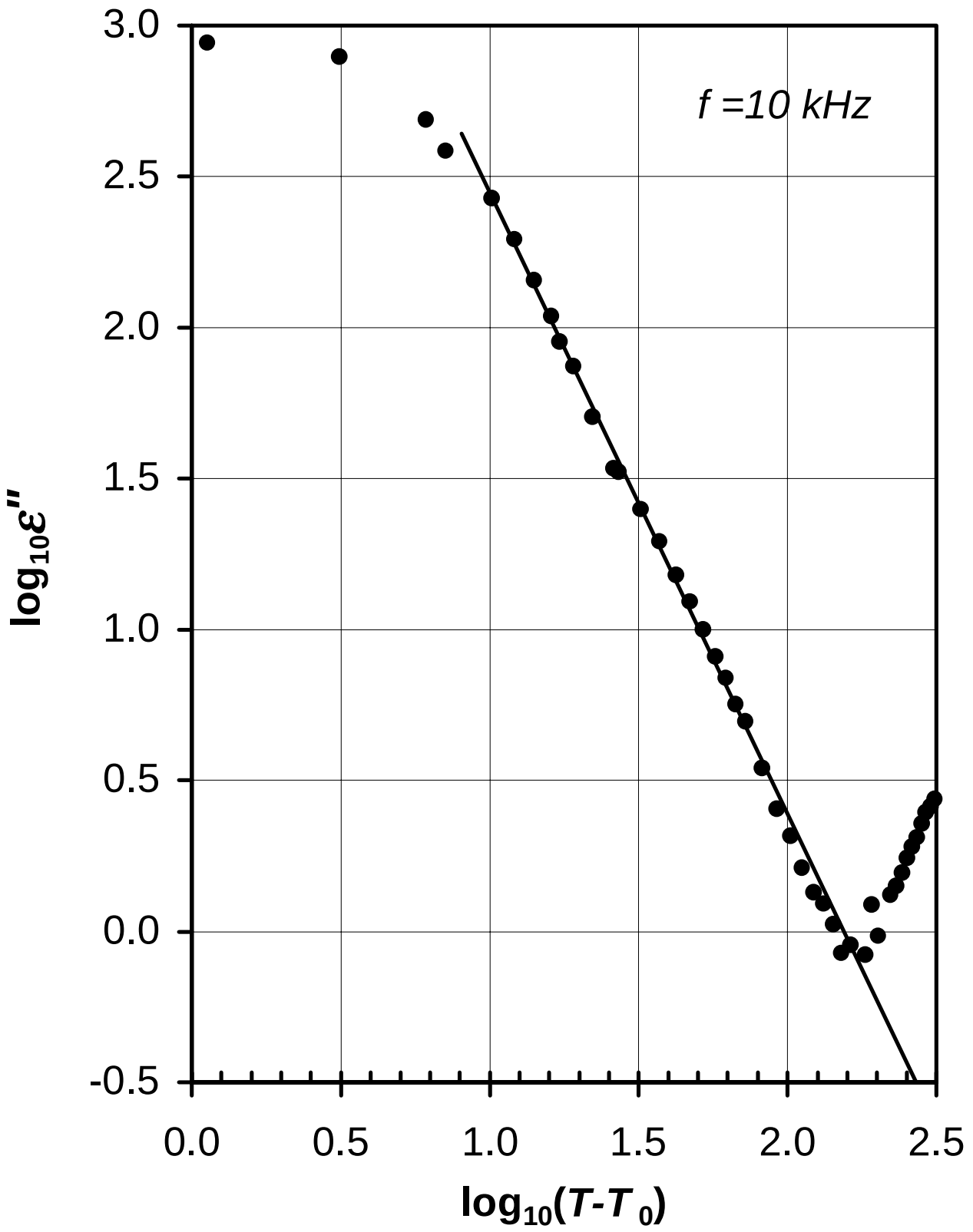

Fig. 8 (Bokov & Ye)

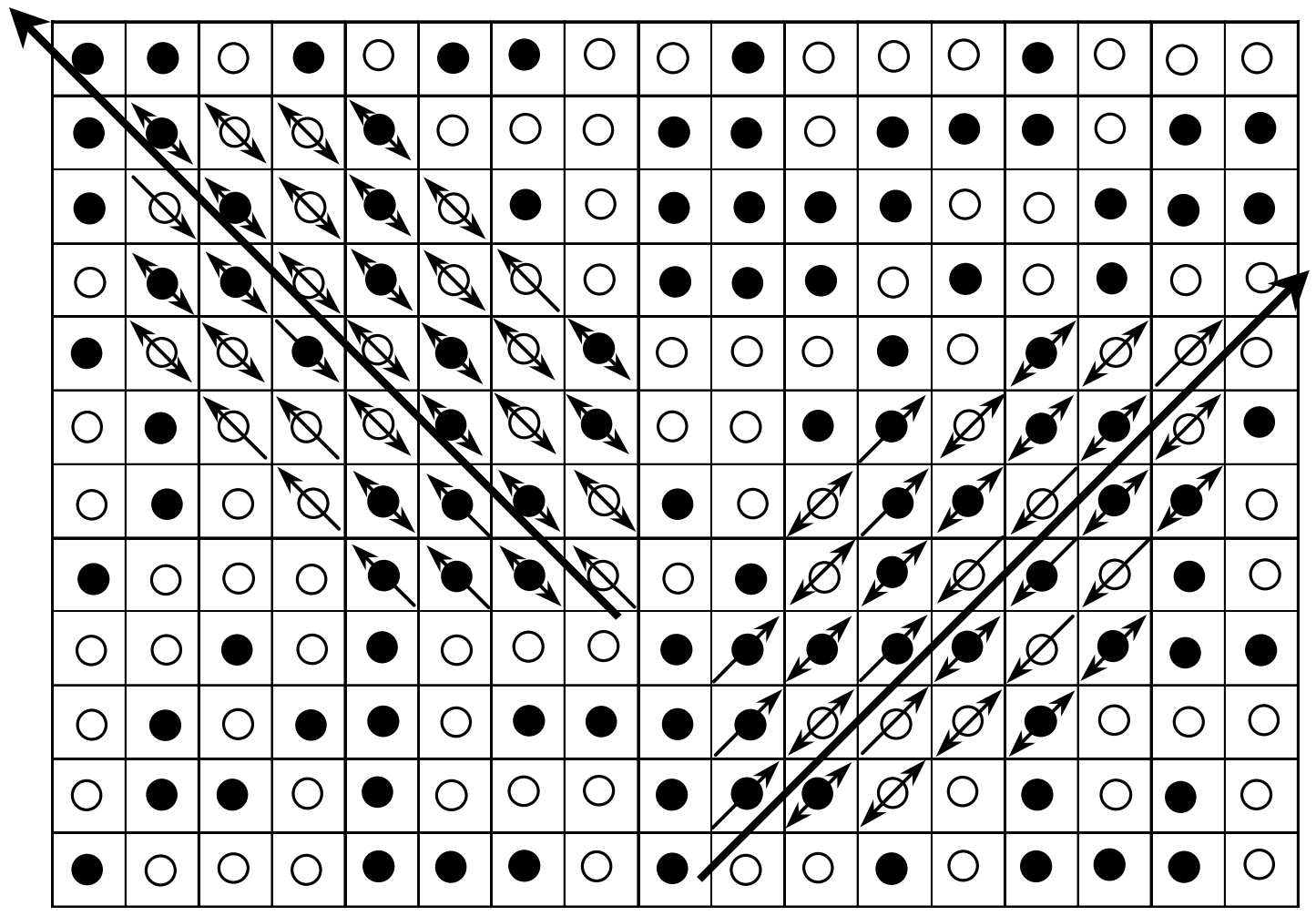

● - B' cations ○ - B" cations

FIG. 9 A.A. Bokov et.al. Phys. Rev.B